\documentclass[10pt]{article}
\usepackage{amssymb}
\usepackage{amsmath}
\usepackage{multirow}
\usepackage{algorithm}
\usepackage{algorithmic}
\usepackage{epsfig}
\usepackage{url}

\newenvironment{varalgorithm}[1]
  {\algorithm\renewcommand{\thealgorithm}{#1}}
  {\endalgorithm}

\newcommand{\makefxn}[1]{\texttt{#1()}}

\begin{document}
\title{Analyzing Non-proportional Hazards: Use of the MRH Package}

\renewcommand{\arraystretch}{0.65}
\author{\small Yolanda Hagar and Vanja Dukic \footnote{
Yolanda Hagar  is a postdoctoral researcher in Applied Mathematics, University of Colorado at Boulder.  Vanja Dukic is a Professor in Applied Mathematics, University of Colorado at Boulder.  Correspondence emails: {\tt yolanda.hagar@colorado.edu, vanja.dukic@colorado.edu.}
}}

\date{  }
\maketitle

\thispagestyle{empty}
\baselineskip 12pt

\begin{abstract}
\noindent
In this manuscript we demonstrate the analysis of right-censored survival outcomes using the MRH package in R.  The MRH package implements the multi-resolution hazard (MRH) model (\cite{Bouman,Bouman2,Dukic,Dignam,Yprune}), which is a Polya-tree based, Bayesian semi-parametric method for flexible estimation of the hazard rate and covariate effects.  The package allows for covariates to be included under the proportional and non-proportional hazards assumption, and for robust estimation of the hazard rate in periods of sparsely observed failures via a ``pruning" tool.

\end{abstract}

\section{Introduction}

The hazard rate, defined as $h(t) = \lim_{\Delta \to 0} {P(t \le T < t+\Delta \mid T \ge t)}/{\Delta} = {f(t)}/{S(t)}$ (where $S(t)$ is the survival function for $T$ and $f(t) = -S'(t)$),  can be critical in assessing how the risk of a disease changes over time.  However, it can be difficult to estimate reliably, particularly over the course of a study with few observed failures during the follow up period and when the effects of covariates change over time. (For examples, see \cite{Andersen, sinha97, mullerbook, KalbPrenticeBook}.)

To this end, the MRH package in R (\cite{MRHR, RuserCompare}) has three overarching features: 
\begin{enumerate}
	\item Estimation of the hazard rate and the associated credible intervals (as well as the corresponding survival function and cumulative hazard estimates and credible intervals). 
	\item Joint estimation of the effects of predictors incorporated under the proportional hazards (PH) and non-proportional hazards (NPH) assumptions.
	\item A ``pruning" tool that combines portions of the hazard rate that are similar, providing robust estimates through periods of sparse failures, and allowing for faster computation times.
\end{enumerate}
The underlying statistical approach employed in the MRH package is the multi-resolution hazard (MRH) model, a Bayesian semi-parametric hazard rate estimator previously presented and used in \cite{Bouman, Bouman2,Dukic,Dignam,Yprune,mrhprost,sadm}, and compared to other packages in \cite{RuserCompare}.  This model for survival data is based on the Polya tree methodology, and is flexibly designed for multi-resolution inference capable of accommodating periods of sparse events and varying smoothness. The MRH model accommodates both  proportional and non-proportional effects of predictors over time and also uses a pruning algorithm presented in \cite{Yprune} and \cite{mrhprost}, which performs data-driven ``pre-smoothing'' of the hazard rate by merging time intervals with similar hazard levels.  Pruning has been shown to increase computational efficiency and reduce overall uncertainty in hazard rate estimation  in the presence of periods with smooth hazard rate and low event counts (\cite{Yprune}).  

The following sections of this manuscript are organized as follows: In Section 2 we provide background on the multi-resolution hazard (MRH) model.  In Section 3 we briefly discuss the tongue cancer data set we use to demonstrate the MRH package, and in Section 4 we cover use of the MRH package, including fitting and plotting fitted models, pruning the prior, assessing convergence of the MCMC chains, and the effects adjustment of the prior parameters.  Lastly, we discuss our conclusions in Section 5.

\section{Multi-resolution hazard modeling}

The MRH model is a Bayesian, semi-parametric survival model that produces an estimate of the hazard rate and covariate effects.  The MRH prior is closely related to the Polya tree prior (\cite{ferguson74, polyaLavine}), which is an infinite, recursive, dyadic partitioning of a measurable space $\Omega$. (In practice, this process is terminated at a finite level $M$.)  The MRH prior is a type of Polya tree in that it uses a fixed, pre-specified partition and controls the hazard level within each bin through a multi-resolution parameterization.  

To facilitate the recursive dyadic partition of the multiresolution tree, we assume that $J=2^{M}$. Here, $M$ is an integer, set to achieve the desired time resolution, or through model selection criteria or clinical input (for example, see \cite{Bouman,Dignam}).  The hazard function is parametrized by a set of  hazard increments  $d_{j}, j = 1,\dots, J$. where $d_j$ represents the aggregated hazard rate over the $j^{th}$ time interval, ranging from $(t_{j-1}, t_j)$.  In standard survival analysis notation,  $d_{j} = \int_{t_{j-1}}^{t_{j}} h(s)ds \equiv H(t_{j})-H(t_{j-1})$, where $h(t)$ is the hazard rate at time $t$.  The cumulative hazard, $H$, is equal to the sum of all  $2^M$ hazard increments, which are denoted as $d_{j}, j = 1,\dots2^M$. The model then recursively splits $H$ at different branches via the ``split parameters" $R_{m,p} = H_{m,2p}/H_{m-1,p}, \,\, m = 1,2,\dots,M-1, \,\, p = 0,\dots,2^{m-1}-1$. Here, $H_{m,q}$ is recursively defined as $H_{m,q}\equiv H_{m+1,2q}+H_{m+1,2q+1}$ (with $H_{0,0} \equiv H$, and $q = 0,\dots,2^{m}-1$).  The $R_{m,p}$ split parameters, each between 0 and 1,  guide the shape of the {\it a priori} hazard rate over time.  The complete hazard rate prior specification  is obtained via priors placed on all tree parameters:  a Gamma($a, \lambda$) prior is placed on the cumulative hazard $H$, and Beta prior on each split parameter $R_{m,p}$,  $\mathcal{B}e(2\gamma_{m,p}k^{m}a,2(1-\gamma_{m,p})k^{m}a)$.  This parametrization ensures the self-consistency of the MRH prior at multiple resolutions (\cite{Bouman, Yprune}).

The basic MRH model was extended in \cite{Dukic} into  the hierarchical  multi-resolution (HMRH)  hazard model,  capable of modeling non-proportional hazard rates in different subgroups jointly with other proportional predictor effects. The pruning methodology for combining similar hazard bins was developed in \cite{Yprune} for individual hazard rates, and combined with the HMRH model in \cite{mrhprost}. The  pruning algorithm detects consecutive time intervals  where failure patterns are statistically similar, increasing estimator efficiency and reducing computing time.  The resulting method produces  computationally stable and efficient inference,  even in periods with sparse numbers of failures, as may be the case in studies with long follow-up periods.   

\subsection{MRH likelihood function}  

We denote $T_i$ as the minimum of the observed time to failure or the right-censoring time for subject $i.$   Each subject belongs to one of the $\mathcal{L}$ covariate strata, and within each stratum we employ the proportional hazards assumption such that: $$h_\ell(t \mid X, \vec\beta) = h_{base, \ell}(t)\exp\{\mathbf  X' {\vec \beta}\}.$$ 
Here, $h_\ell$ denotes the baseline hazard rate for treatment strata $\ell$,  ${\mathbf X}$ represents the $z \times n_{\ell}$  matrix of $z$ covariates (other than those used for stratification) for the $n_{\ell}$ patients in the stratum $\ell$, while ${\vec \beta}$ denotes the $z \times 1$ vector of the covariate effects. 

For subject $i$ in stratum $\ell$ with failure time at $T_{i} \in [0, t_J)$, the likelihood contribution is: 
\begin{align*}	
	L_{i, \ell}(T_{i} \mid X_{i}, \vec \beta) = h_{base, \ell}(T_{i})\exp(X_{i}'\vec \beta)^{\delta_i}S_{base, \ell}(T_{i})^{\exp(X_{i}'\vec \beta)},
\end{align*}
where $X_{i}$ is that subject's covariate vector, $S_{base, \ell}$ is the baseline survival function for the stratum $\ell$, and $\delta_{i}$ is the censoring indicator that equals 1 if subject $i$ had an observed failure, and 0 otherwise.  Thus, the log-likelihood for all $n$ patients in all $\mathcal{L}$ strata together ($n = \sum_{\ell = 1} ^{\mathcal{L}} n_{\ell}$) is
\begin{eqnarray*}
	\log L(\mathbf{T \mid \vec \beta, H, R_{m,p}, X}) = \sum_{\ell = 1}^{\mathcal{L}}\sum_{i \in \mathcal{S}_\ell}\left\{
			\delta_i\log\left(h_{base, \ell}(T_i) + X_i'\vec \beta\right) - \exp\left(X_{i}'\vec \beta\right)H_{base, \ell}(T_i)\right\},
\end{eqnarray*}
where ${\mathcal{S}_\ell}$ denotes the set of indices for subjects belonging to the stratum $\ell$,  and $H_{base, \ell}(T) = -\log S_{base, \ell}(T).$  In this model, the $\mathcal{L}$ hazard rates are estimated jointly with all the covariate effects.  The non-proportional covariate effect is then calculated as the log of the hazard ratio between different covariate strata in each bin.  

\subsection{Pruning the MRH model} 
The  MRH prior resolution is often chosen as a compromise between the desire for detail in the hazard rate, and the number of observed (i.e. uncensored) failures. As  the resolution increases, the number of observed failures within each bin decreases. While useful for revealing detailed patterns, a large number of  intervals is a large number of model parameters, which will generally require longer computing times and may result in estimators with lower statistical efficiency (\cite{Yprune}). 

``Pruning''  starts with the full MRH tree prior, and merges adjacent bins that are constructed via the same split parameter, $R_{m,p}$,  when the hazard increments in these two bins ($H_{m+1,2p}$ and $H_{m+1,2p+1}$) are statistically similar.  This is inferred by testing the hypothesis $H_0: R_{m,p}=0.5$  against the alternative $H_a: R_{m,p} \neq 0.5$, with a pre-set type I error $\alpha$, using Fisher's exact test. If the null hypothesis is not rejected, that split $R_{m,p}$ is set to $0.5$ and the adjacent hazard increments are considered equal and the time bins declared ``fused''.  The hypothesis testing can be applied  to all $M$ levels of the tree or just a higher resolution subset of the tree.  Because bins are fused \textit{a priori}, this method can then reduce the number of parameters sampled in the MCMC routine, possibly decreasing computation time and increasing the robustness of the estimator.  

\subsection{Estimation in the MRH model}
Estimation is performed in two steps: the pruning step and the MCMC steps.  The pruning step is run only once for each of the $\mathcal{L}$ hazard rates at the beginning of the algorithm as a pre-processing step in order to finalize the MRH tree priors. The $R_{m,p; \ell}$ parameters for which the null hypothesis is not rejected are set to $0.5$ with probability 1, while the rest are estimated in the Markov chain Monte Carlo (MCMC) routine.  Details on the MCMC routine and prior values can be found in \cite{Bouman, Dukic} and \cite{mrhprost}. 

\section{Tongue cancer data}

To demonstrate the different features of the package, we use the ``tongue" data set available in the R data set available in the R survival package. The data set contains 80 subjects with tongue cancer who had a paraffin-embedded sample of the cancerous tissue taken at the time of surgery, with survival times recorded for each patient (in weeks), as well as the tumor DNA profile (aneuploid or diploid) (see \cite{SickleSant} for details).  The study went for 400 weeks, with a median survival time equal to 69.5 weeks (SD = 67.3), and 33.8\% of the subjects were censored.  Between 250 and 350 weeks there were zero failures (censored or uncensored), and after 200 weeks there were zero uncensored failures. This data set is also presented in \cite{KleinandMo} and analyzed \cite{EucRef}. Table \ref{tab:summcateg} summarizes the data in more detail.
\begin{table}
	\caption{Sample characteristics of 80 patients in the ``tongue" data set found in the R survival package, stratified by tumor type.}\label{tab:summcateg}
\centering
\begin{tabular}{|c|cc|cc|cc|}
	\hline
	&&&&&&\\[.01ex]
	& \multicolumn{2}{|c|}{Aneuploid} & \multicolumn{2}{|c|}{Diploid} &\multicolumn{2}{|c|}{Total Sample}\\[.01ex]
	\hline
   	& N& \% & N&  \%  & N&  \%      \\[.01ex]
	\hline
	\hline
	&&&&&&\\[.01ex]
	Uncensored &31&59.6&22&78.6&53&66.25\\[.5ex]
	\hline
	\hline
	&&&&&&\\[.01ex]
	Censored&21&40.4&6&21.4&27&33.75\\
	\hline
	\hline
	&&&&&&\\[.01ex]
	Total&52&65.0&28&35.0&80&100.0\\[.5ex]
	\hline
\end{tabular}
\end{table}

In the code shown in Section \ref{sec:analysis}, the tongue data is named ``tongue", and has the following variable names:
\begin{itemize}
	\item time: Survival time (in weeks) from time of surgery.
	\item type: Indicator denoting which tumor group the patient belongs to (`1' = aneuploid, `2' = diploid).
	\item delta: The censoring indicator, which equals `1' if the failure is observed, and `0' if right-censored.
\end{itemize}
\section{Using MRH}\label{sec:analysis}
In this section, we provide code and discussion on how to analyze survival data using the MRH package, demonstrating analyses using the tongue data to quantify survival times post-surgery.

\subsection{Selecting the time resolution (specifying $M$)}
Because the MRH methodology is based on a binary partition, it divides the total study time into $J = 2^M$ time intervals (or ``bins"). The first step to fitting an MRH model in R is to determine how many bins are required. (The current version of the MRH package assumes all bins are of equal length.)   The choice of $M$ is can be determined through biological rationale or using penalized likelihood criteria (such as DIC, see \cite{refDIC}).  
In some cases the user might wish to explore how different choices of $M$ affect the bin lengths and implications on the model interpretation.  In these instances, the \makefxn{FindBinWidth} function can help.  Below, the bin width is calculated for values of $M$ ranging from 2 to 10 for different time units (seconds, minutes, hours, days, weeks, months, years).  The user must provide the vector of survival times, and specify the original unit of the survival times (seconds (`s'), minutes (`m'), hours (`h'), days (`d'), weeks (`w'), months (`m'), and years (`y')).   
\begin{verbatim}
data(tongue)

 FindBinWidth(time = tongue$time, delta = tongue$delta, time.unit = 'w')
[1] The mean failure time is  73.825 weeks
[1] The median failure time is  69.5 weeks
[1] The range of failure times is  1 to 400 weeks
 
        secs      mins     hours       days      weeks      months       years
M2  60480000 1008000.0 16800.000 700.000000 100.000000 22.99794661 1.916495551
M3  30240000  504000.0  8400.000 350.000000  50.000000 11.49897331 0.958247775
M4  15120000  252000.0  4200.000 175.000000  25.000000  5.74948665 0.479123888
M5   7560000  126000.0  2100.000  87.500000  12.500000  2.87474333 0.239561944
M6   3780000   63000.0  1050.000  43.750000   6.250000  1.43737166 0.119780972
M7   1890000   31500.0   525.000  21.875000   3.125000  0.71868583 0.059890486
M8    945000   15750.0   262.500  10.937500   1.562500  0.35934292 0.029945243
M9    472500    7875.0   131.250   5.468750   0.781250  0.17967146 0.014972621
M10   236250    3937.5    65.625   2.734375   0.390625  0.08983573 0.007486311
\end{verbatim}
While there are many acceptable bin widths for an analysis, we generally aim to use one that will allow us to observe a maximum amount of detail in the hazard rate while still remaining computationally feasible and biologically plausible.  In a typical analysis, we would be inclined to use a model with $M = 6$, creating bins that are 6.25 weeks long.  However, for the purposes of this manuscript, we reduce the number of bins to 16 (i.e. $M = 4$) to reduce computing time while still allowing for adequate demonstration of the package.

\subsection{Fitting the MRH model} 

Once $M$ has been determined, the MRH model is fit using \makefxn{estimateMRH}.  In the examples below, $M$ is equal to 4, with each bin representing 25 weeks. We examine MRH models including the treatment covariate under the proportional and non-proportional hazards assumptions, as well as with various levels of pruning.  

\subsubsection{Proportional hazards model}\label{sec:PH}
In this example, we include the tumor-type in the model under the proportional hazards assumption.  The code to fit the model is below: 
\begin{verbatim}
fit.PH = estimateMRH(Surv(time, delta) ~ type, data = tongue,
	M = 4, maxStudyTime = 400, outfolder = 'MRHresults_PH')
 
\end{verbatim}
The user is notified of the approximate running time and after every 5,000 iterations completed by the routine.  The output for each model is placed into a sub-folder created by the \makefxn{estimateMRH} routine within the working directory.  The default folder is named ``MRHresults", however, in this example we have specified the output folder as ``MRHresults\_PH" through the ``outfolder" option. (The folder name can also be added to a pathname, but the path must be accessible from the working directory.)  Because parameters are estimated through MCMC sampling, convergence of the fitted model should be assessed after the routine has finished (see Section \ref{sec:MCMCconvergence} for details).

\begin{figure}[h!]
	\centering
	\includegraphics[width = 4.5in]{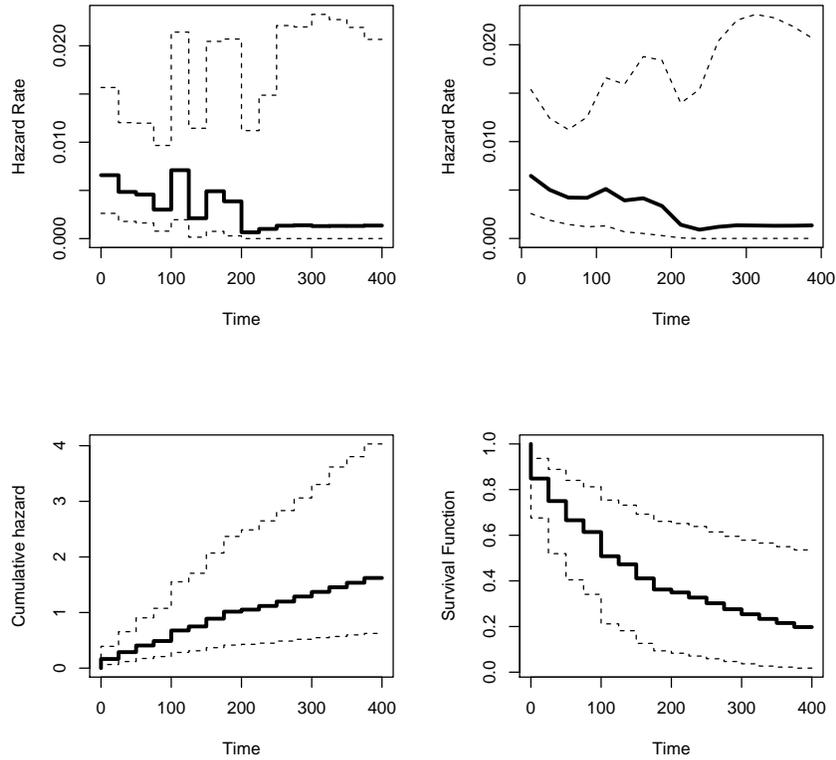}
	\caption{\footnotesize Four graphs of the fitted PH model ``fit.PH", all created using the \makefxn{plot.MRH} function and associated options on the MRH fitted object.  TOP LEFT: Default graph of the hazard rate of death from tongue cancer (\texttt{plot(fit.PH)}).  TOP RIGHT: Smoothed graph of the hazard rate of death from tongue cancer (\texttt{plot(fit.PH, smooth.graph = TRUE, smooth.df = 10)}).  BOTTOM LEFT: The cumulative hazard (\texttt{plot(fit.PH, plot.type = 'H')}).  BOTTOM RIGHT: The survival function (\texttt{plot(fit.PH, plot.type = 'S')}).}
	\label{fig:PHautographs}
\end{figure}

The fitted model can be examined using both the \makefxn{summary.MRH} function as well as the \makefxn{plot.MRH} function (see Figure \ref{fig:PHautographs} for plot results):
\begin{verbatim}
# Get the estimated model results
results = summary(fit.PH)
names(results)
[1] "hazardRate"       "beta"             "SurvivalCurve"    "CumulativeHazard"
[5] "d"                "H"                "Rmp"          
results$hazardRate
            hrEst    hrq.025    hrq.975
h.bin1  0.00657792 0.00261944 0.01567524
h.bin2  0.00484148 0.00178080 0.01202800
h.bin3  0.00457100 0.00162196 0.01197604
h.bin4  0.00300732 0.00077372 0.00966172
h.bin5  0.00709884 0.00195920 0.02141968
...
# Plot the results (default, smoothed version, cumulative hazard, survival curve)
plot(fit.PH)
plot(fit.PH, smooth.graph = TRUE, smooth.df = 10)
plot(fit.PH, plot.type = 'H')
plot(fit.PH, plot.type = 'S')
\end{verbatim}
(Note that in this manuscript, we may use `...' when showing output to indicate that further output is available, but not shown for the legibility purposes.)

\subsubsection{Non-proportional hazards model}
\begin{figure}
	\centering
	\includegraphics[width = 4.5in]{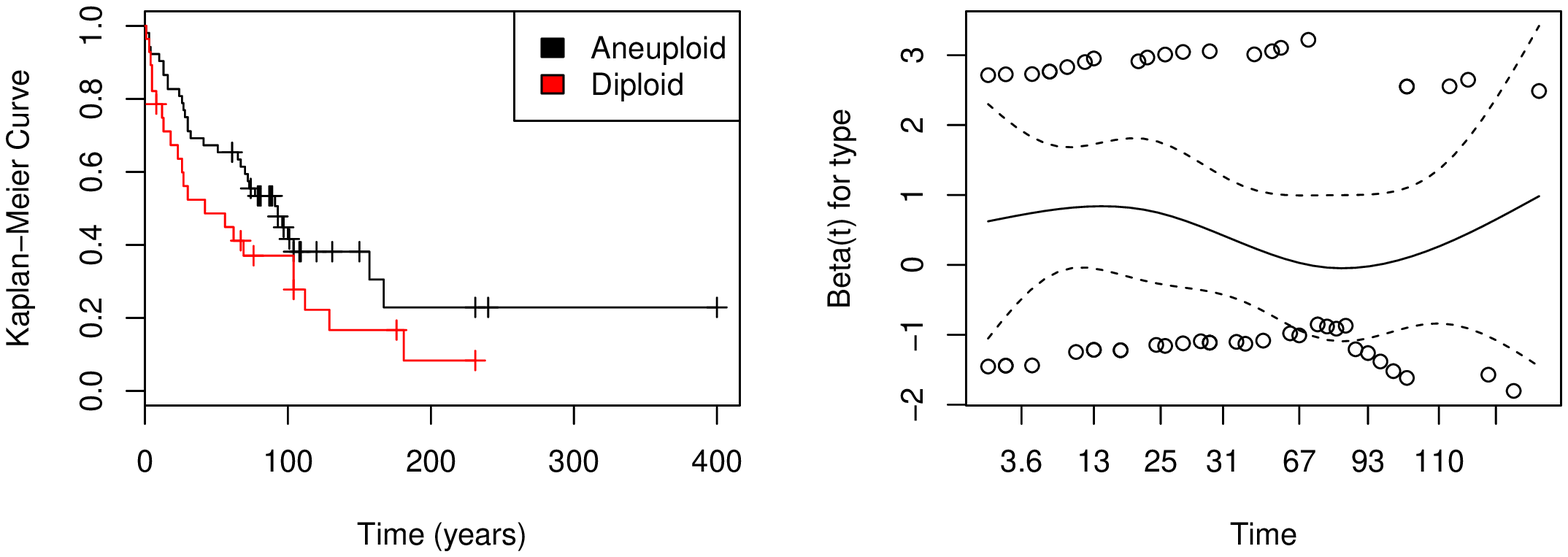}
	\caption{\footnotesize LEFT: Kaplan Meier survival curves of the two treatment groups.  The two survival curves do not appear proportional to one another, particularly as they get closer in value towards the end of the study.  In addition, each `+' sign indicates a censored observation, so many subjects do not have observed biochemical failure.  RIGHT: A plot of the Schoenfeld residuals (based on work done by \cite{SchoenRes}) shows a smoothed line that does not appear to be straight, indicating non-proportionality between the hazards.  The information provided by these two graphics warrants analyses including the treatment covariate under a non-proportional hazards assumption.}
	\label{fig:NPHcurves}
\end{figure}

In many studies with long-term follow-up, the effects of certain covariates change over time, violating the proportional hazards assumption.  A figure of the Kaplan-Meier curves (\cite{KMest}) (see Figure \ref{fig:NPHcurves}, left) show evidence that the effects of treatment may not be proportional.  In addition, we can examine if the treatment effect seems to be proportional by creating a cox model (using \makefxn{coxph} in the survival package) and then examine the Schoenfeld residuals and smoothed line (see \cite{SchoenRes}), which should be straight if the effects of tumor type were proportional (see Figure \ref{fig:NPHcurves}, right).  Results from these figures provide evidence that the treatment effect is not proportional.

Because of this evidence of non-proportionality between the two treatment group hazard rates, we modify the model above using the \makefxn{nph} function in the formula portion used to fit the model:

\begin{verbatim}
fit.NPH = estimateMRH(Surv(time, delta) ~ nph(type), data = tongue,
	M = 4, maxStudyTime = 400, outfolder = `MRHresults_NPH')
\end{verbatim}

In creating a model with at least one NPH covariate, it is important to note that  the non-proportional covariate must be a nominal categorical variable.  More than one non-proportional covariate can be entered in to the model, using repeated \makefxn{nph} functions in the formula, and the routine will merge the NPH variables into a single interaction variable, jointly estimating stratified hazard rates for each combination of levels.  See \texttt{vignette("MRH")} for more information.

Summaries for the NPH fitted model are displayed using \makefxn{summary.MRH}, with separate estimates provided for each hazard rate, as well as the estimated log-ratio between the hazard rates:
\begin{verbatim}
summary(fit.NPH)$hazardRate
                 hrEst hrq.025 hrq.975
                    hrEst    hrq.025    hrq.975
h.bin1.group1  0.00824948 0.00415240 0.01457768
h.bin2.group1  0.00727956 0.00315452 0.01440876
h.bin3.group1  0.00694756 0.00281784 0.01407212
h.bin4.group1  0.00718136 0.00223352 0.01665544
h.bin5.group1  0.00716788 0.00138364 0.02265908
...
h.bin12.group2 0.00713068 0.00002044 0.16397088
h.bin13.group2 0.00720296 0.00001700 0.16742536
h.bin14.group2 0.00703252 0.00001728 0.16919512
h.bin15.group2 0.00759180 0.00002364 0.18139816
h.bin16.group2 0.00706220 0.00001868 0.17366552

summary(fit.NPH)$beta
                    betaEst betaq.025 betaq.975
beta.type.2.bin1   0.810746 -0.112749  1.701116
beta.type.2.bin2   0.403418 -0.910105  1.572686
beta.type.2.bin3   0.367961 -1.117286  1.688406
beta.type.2.bin4  -2.127125 -6.461301  0.686255
...
\end{verbatim}

\begin{figure}[h!]
	\centering
	\includegraphics[width = 2.25in]{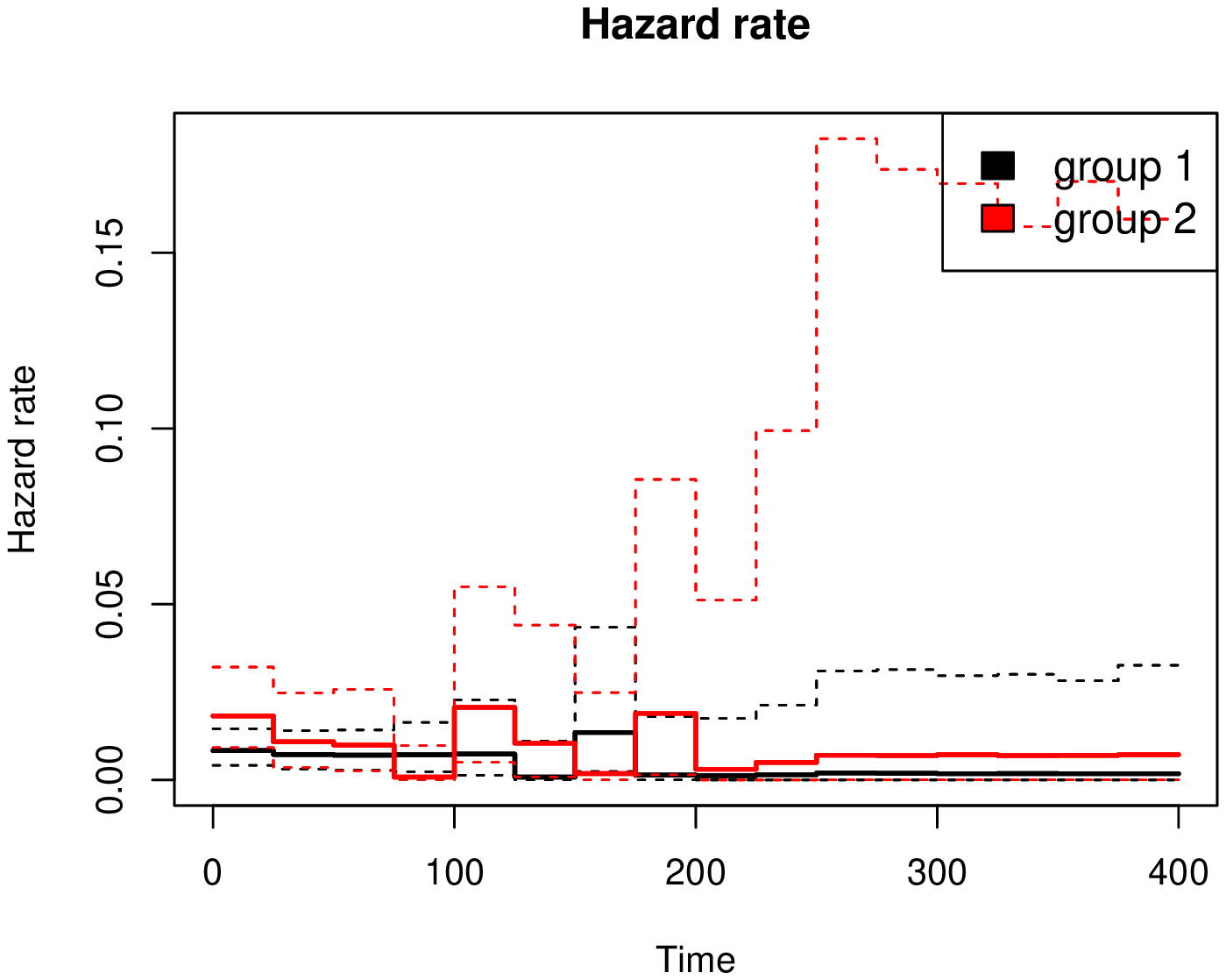}
	\includegraphics[width = 2.25in]{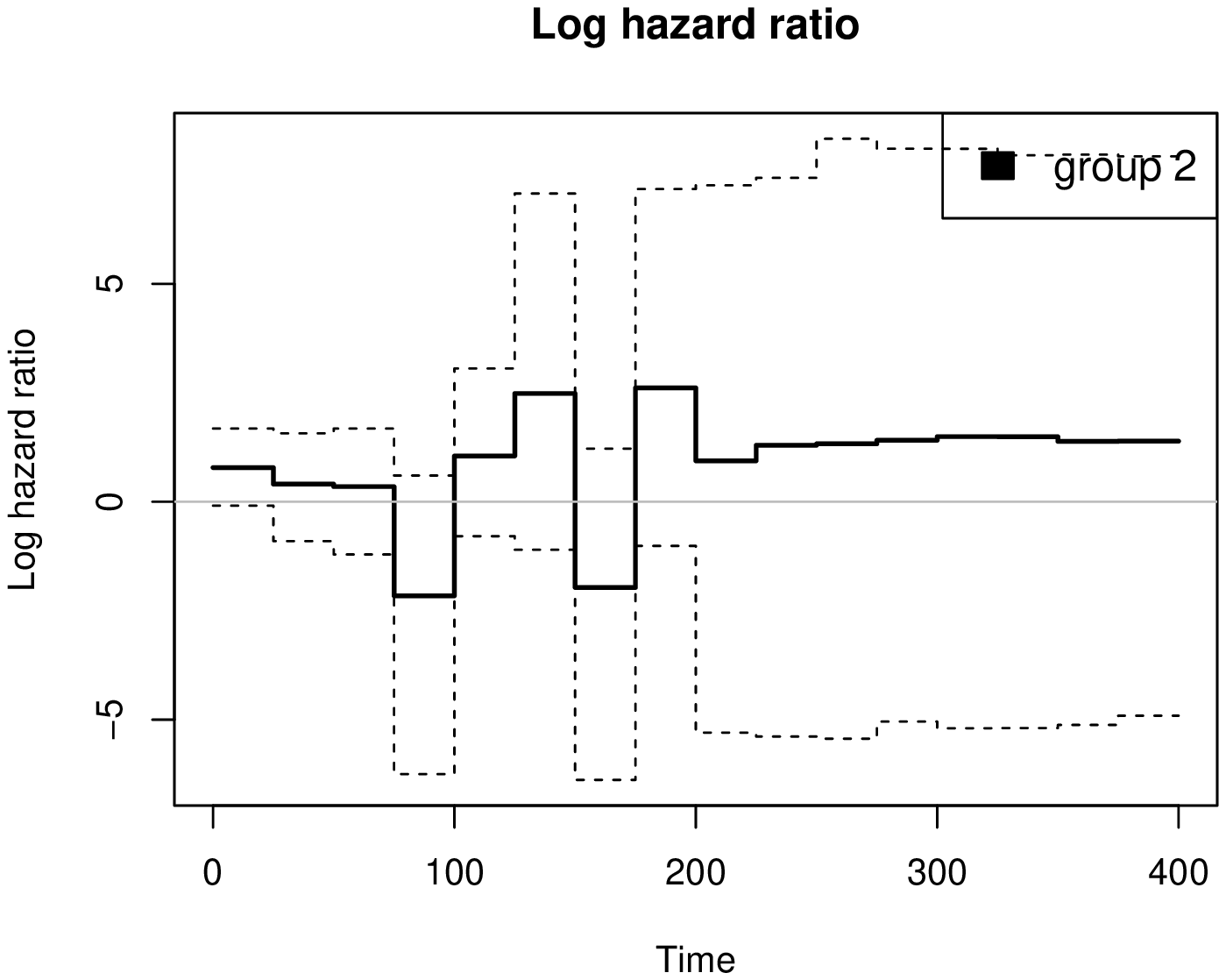}
	\includegraphics[width = 4in]{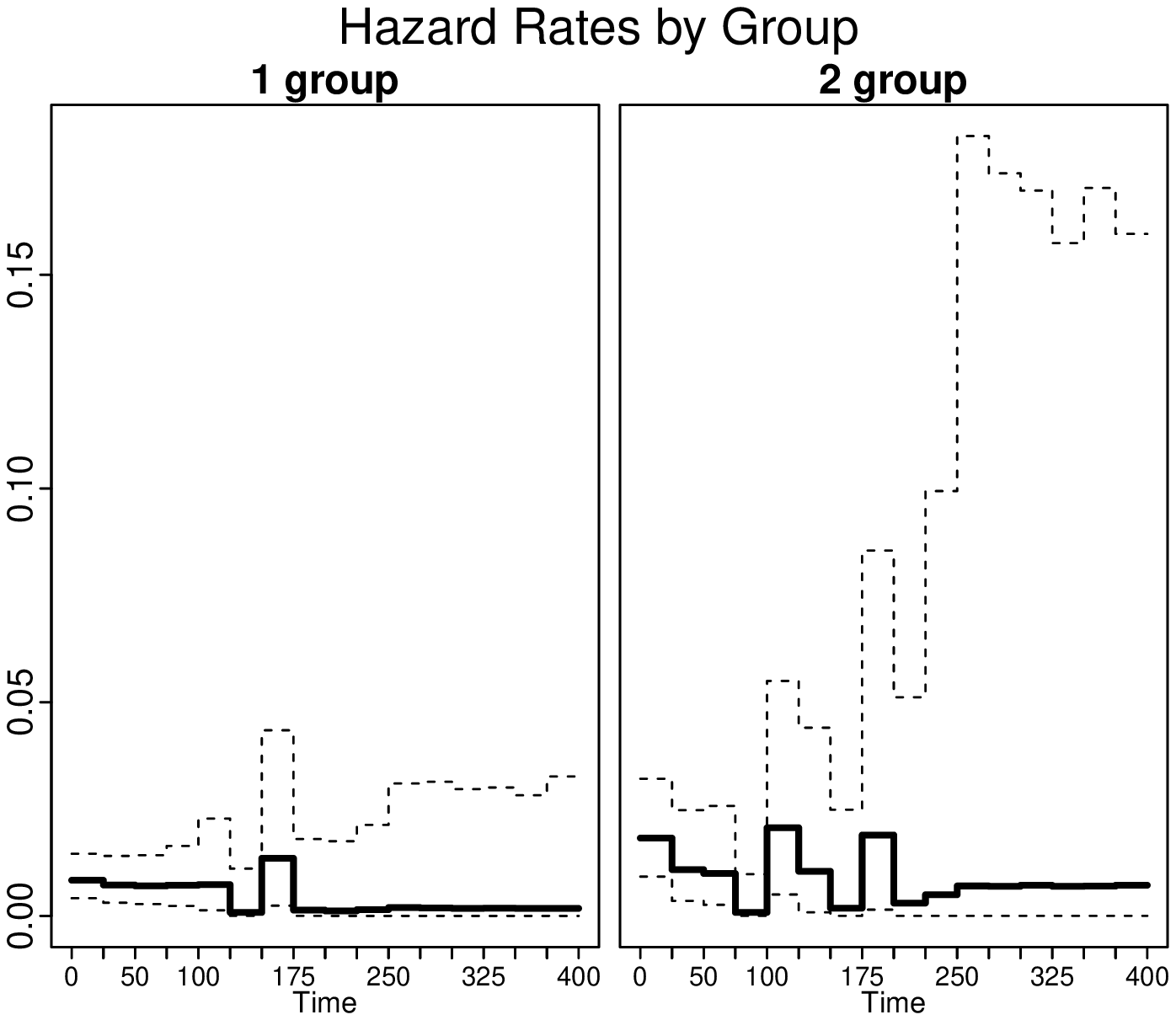}
	\caption{\footnotesize Three graphs of the fitted NPH model, all created using the \makefxn{plot.MRH} function and associated options on the MRH fitted object.  TOP LEFT: Default graph of the hazard rates of biochemical failure for each treatment group (\texttt{plot(fit.NPH)}).  TOP RIGHT: Log-hazard ratio of the treatment effect over time (\texttt{plot(fit.NPH, plot.type = 'r')}).  BOTTOM: The estimated hazard rates for each treatment group, each shown on a separate graph (\texttt{plot(fit.NPH, combine.graphs = FALSE)}).}
	\label{fig:NPHautographs}
\end{figure}

Plots of the estimated hazard rates and the log-hazard ratio of the NPH covariate can also be created using \makefxn{plot.MRH}.  Example graphing code is below (see Figure \ref{fig:NPHautographs} for plot results): 
\begin{verbatim}
# Plot the default graph (the hazard rate of each treatment group)
plot(fit.NPH)
# Plot the log-hazard ratio of the treatment effect
plot(fit.NPH, plot.type = 'r')
# Plot the hazard rates each on a separate graph
plot(fit.NPH, combine.graphs = FALSE)
\end{verbatim}
Note that the plotting options shown in Section \ref{sec:PH} can also be used in the NPH plots.  Details on convergence checking, output generation, and other specifications on the MCMC chains are found in Section \ref{sec:MCMCconvergence}.  

\subsection{Pruning the MRH model}
As mentioned previously, in instances where the number of observed failures is small, estimates of the hazard rate can be difficult to obtain due to lack of information.  In the examination of biochemical failure, the number of subjects who have observed biochemical failure is small, particularly towards the end of the study where a lot of censoring occurs.  In this particular data set, only 50\% of subjects have an observed biochemical failure, and only 13\% of the observed failures occur after 5 years.

In the MRH package, bins can combined using the ``pruning" method (\cite{Yprune}), implemented using the ``prune" option in the \makefxn{estimateMRH} function. It is possible to prune one to all levels of the prior tree based on biological rationale and the degree of possible smoothing the user desires.  By default, all levels of the tree are pruned using a significance level $\alpha = 0.05$.  However, the number of levels pruned can be adjusted using the ``prune.levels" option, and the significance level can be controlled using the ``prune.alpha"  option.  (Note that smaller values of $\alpha$ will lead to smoother prior trees, as the null hypothesis that the two bins are similar will not be rejected as often.) 

\begin{figure}[h!]
	\centering
	\includegraphics[width = 4.5in]{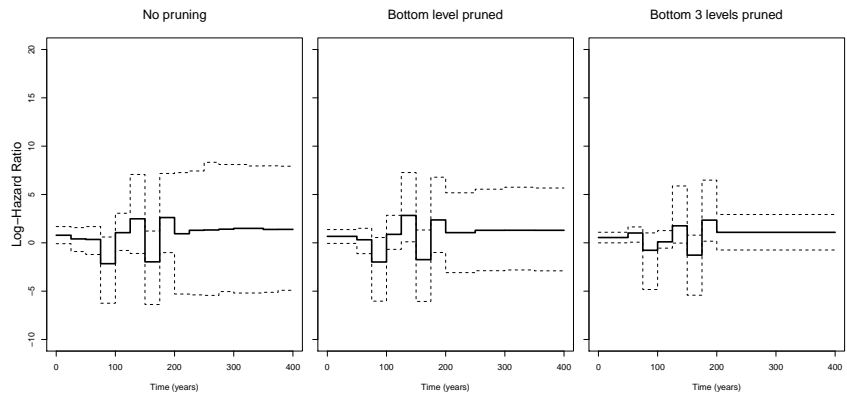}
	\caption{\footnotesize Graph comparing the estimated log-hazard ratio (solid black line) and 95\% credible intervals (grey shading) for the three NPH models, each calculated using different levels of pruning. On the left, the model with no pruning shows wider error bounds than the 3-level pruned model (center) and then all-level pruned model (right).}
	\label{fig:compareprune}
\end{figure}

Below, we show code for pruning the MRH prior tree for the NPH model, combining bins in the bottom 3 levels and combining bins in all levels.  In the NPH model, each hazard rate is pruned separately:

\begin{verbatim}
# Model with no pruning (same as the model shown in Section 4.2)
fit.NPH = estimateMRH(Surv(time, delta) ~ nph(type), data = tongue,
      M = 4, maxStudyTime = 400, outfolder = 'MRHresults_NPH')

# Model with the bottom level of the tree pruned
fit.NPHprune1 = estimateMRH(Surv(time, delta) ~ nph(type), data = tongue,
      M = 4, maxStudyTime = 400, outfolder = 'MRHresults_NPHprune1',
      prune = TRUE, prune.levels = 1)

# Model with bottom 3 levels of the tree pruned
fit.NPHprune3 = estimateMRH(Surv(time, delta) ~ nph(type), data = tongue,
      M = 4, maxStudyTime = 400, outfolder = 'MRHresults_NPHprune3',
      prune = TRUE, prune.levels = 3)

\end{verbatim}
Results of the estimates for the three different models can be seen in Figure \ref{fig:compareprune}.  As expected, the model with no pruning (left graph) has the highest variability among the three models.  The model with the bottom 3 levels pruned shows more variability than the model with all levels pruned, however few bins were merged in the top 3 levels of the tree, so results for the two models (shown in the center and right graphs) are similar.
\subsection{Prior parameter effects}
In the \makefxn{estimateMRH} function, the user has the option to adjust the parameters of the prior distribution for the split parameters.  The prior for the $R_{m,p}$ parameters is a beta distribution, with the form:
$$R_{m,p} \sim Beta\left(2\gamma_{m,p}k^ma, 2(1-\gamma_{m,p})k^ma\right).$$

It is possible to sample the parameters $k$ and/or $\gamma$ using the \makefxn{estimateMRH} routine.  Alternatively, specific fixed values of $k$ and $\gamma$ can be defined {\it a-priori} to influence the shape and smoothness of the hazard rate (the default is fixed values equal to 0.5 for all parameters). Details on the role of these hyperparameters can be found in \cite{Bouman,Bouman2,Yprune,mrhprost}. Below is a brief explanation of the assumptions that stem from using different fixed or sampled values of $k$ and $\gamma.$  

\subsubsection{Comparison and selection of $k$}
In the MRH package, by default $k$ is fixed 0.5, implying zero {\it a-priori} correlation among the hazard increments.  However, $k$ can be fixed with values greater than 0.5, implying the increments are positively correlated \textit{a priori} (typically leading to smoother estimated hazard functions).  Alternatively, if $k$ is fixed with a value less than 0.5, the hazard increments are negatively correlated \textit{a priori} and generally cause estimated hazard rates that are less smooth.  It is also possible to sample the parameter $k$ if the user desires. The user may change the default value of $k = 0.5$ by entering a value for``k.fixed," with $k \in (0, \infty),$ in the {\tt estimateMRH} function.  Below, we show example code for sampling $k$, and fixing $k$ at 0.1 and 10:
\begin{verbatim}
# Sample the k parameters for each hazard rate
fit.NPH.samplek = estimateMRH(Surv(time, delta) ~ nph(type), data = tongue,
      M = 4, maxStudyTime = 400, outfolder = 'MRHresults_NPH_ksample',
      k.fixed = FALSE)

# Fix k at 0.1 
fit.NPH.kpt1 =estimateMRH(Surv(time, delta) ~ nph(type), data = tongue,
      M = 4, maxStudyTime = 400, outfolder = 'MRHresults_NPH_kpt1',
      k.fixed = .1)

# Fix k at 10
fit.NPH.k10 = estimateMRH(Surv(time, delta) ~ nph(type), data = tongue,
      M = 4, maxStudyTime = 400, outfolder = 'MRHresults_NPH_k10',
      k.fixed = 10)
\end{verbatim}

\begin{figure}[h!]
	\centering
	\includegraphics[width = 4in]{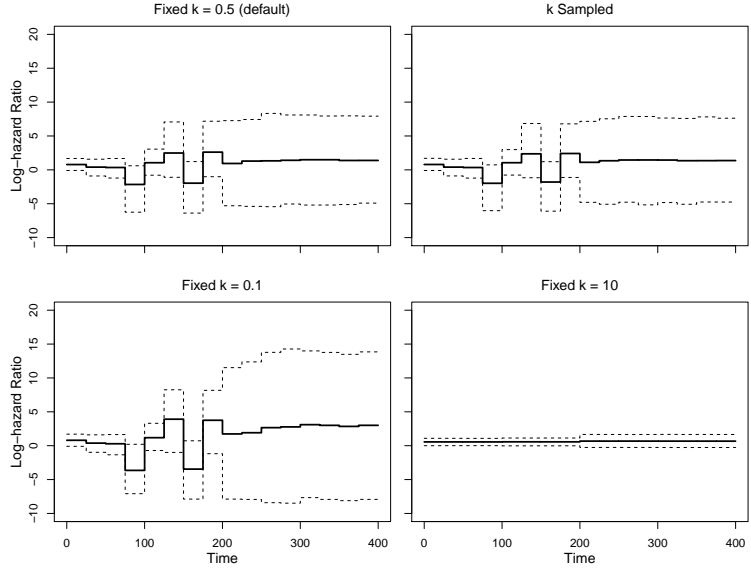}
	\caption{\footnotesize Comparison of the estimated log-hazard ratio of the treatment effect for the NPH model, with $k$ included in the $R_{m,p}$ prior in different ways: fixed at 0.05, implying zero \textit{a priori} correlation between bins (the default value, top left), $k$ sampled (top right), $k$ fixed at 0.1, implying negative correlation between bins (bottom left) and $k$ fixed at 10, implying positive correlation between bins (bottom right).  Estimates are shown with solid lines, and 95\% credible intervals are shown with dashed lines.  As expected, the model using the smallest value of $k$ produces the estimate with the most peaks (i.e. the ``bumpiest" estimate), and the model using the largest value of $k$ (equal to 10) produces the smoothes estimate.  The model that samples $k$ produces an estimate very close to the default model, with median estimates of $k$ equal to 0.55 for the aneuploid group and 0.56 for the diploid group.} 
	\label{fig:comparek}
\end{figure}

Note that in the NPH models, a vector of $k$ values can be entered, with one $k$ for each subgroup hazard rate.  However, if only one value of $k$ is specified, that value will be used for all hazard rates.  Alternatively, if ``k.fixed" is set to \texttt{FALSE}, the routine will sample the $k$ parameter(s), putting an exponential hyper prior on $k$.  Estimated log-hazard ratios of the treatment covariate from the models coded above can be observed in Figure \ref{fig:comparek}.  As anticipated, the estimate from the model where $k$ is sampled (upper right) has the highest variability.  In contrast, the model with a very large $k$ value (lower right) has the lowest variability and also a very smooth estimated function.  

\subsubsection[Comparison and selection of gammamp]{Comparison and selection of $\gamma_{mp}$}

The mean of the prior for each split parameter $R_{m,p}$ is $E(R_{m,p}) = \gamma_{m,p},$ with $\gamma_{m,p} \in (0,1).$ This allows the user to \textit{a priori} ``center" the baseline hazard increments in each bin at a desired value (please see \cite{Bouman2} for details).  

\begin{figure}[h!]
	\centering
	\includegraphics[width = 4in]{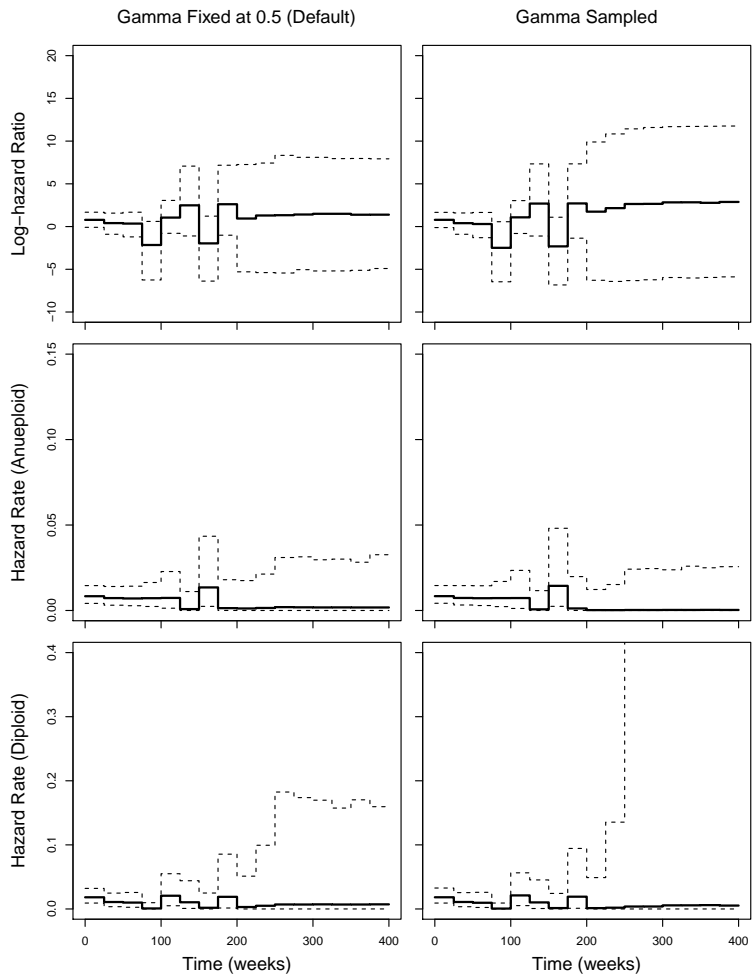}
	\caption{\footnotesize Comparison of the estimated log-hazard ratio and hazard rates (solid lines), and 95\% credible intervals (dashed lines) for both treatment groups for the default MRH model (left column) and the MRH modeled with sampled $\vec\gamma$ values (right column).  While the estimated hazard rates and log-hazard ratios do not differ dramatically, the variation in the model with sampled $\vec\gamma$ values is greater.  This is expected, as the second model estimates $2^{M-1}\times2$ more parameters than the default model.} 
	\label{fig:samplegamma}
\end{figure}

The default value for the $\gamma$ parameters is 0.5, however the user may change the default values via the ``gamma.fixed" option.  In general, unless the user has specific information that would help in the adjustment of the $\gamma$ parameters, we recommend keeping the value fixed at 0.5. In the case of user input, a $\gamma_{m,p}$ value must be specified for each of the split parameters.  In NPH models, a matrix of fixed $\gamma$ values may be entered, with each column representing the desired values for each subgroup hazard rate.  However, this is not required; if only a vector is specified, that vector will be used for all hazard rates.  Below, we show code for sampling the $\gamma$ parameters, putting a beta hyper prior on each $\gamma_{m,p}$:
\begin{verbatim}
fit.NPH.gammasample = estimateMRH(Surv(time, delta) ~ nph(type), data = tongue,
      M = 4, maxStudyTime = 400, outfolder = 'MRHresults_NPH_gsample',
      gamma.fixed = FALSE)

summary(fit.NPH.gammasample)$gamma
             gammampEst gammampq.025 gammampq.975
gammamp1.0_1   0.755960     0.275969     0.972982
gammamp2.0_1   0.513516     0.134032     0.884236
gammamp2.1_1   0.461246     0.055678     0.935269
gammamp3.0_1   0.512232     0.113273     0.890278
gammamp3.1_1   0.451540     0.080873     0.877487
...
gammamp4.3_2   0.375792     0.057736     0.854770
gammamp4.4_2   0.478627     0.077879     0.913257
gammamp4.5_2   0.499923     0.077736     0.924315
gammamp4.6_2   0.497371     0.092031     0.917733
gammamp4.7_2   0.502462     0.085781     0.923276
\end{verbatim}
The estimated hazard rates and log-hazard ratio can be observed in Figure \ref{fig:samplegamma}.  The estimates do not differ that dramatically, but the variation in the model where the vector of $\gamma$ parameters is sampled is greater.  This is expected, since this model estimates $2^{M-1}\times2$ more parameters than the default model.

\subsection{Model comparison}
In comparing different models, it may be desirable to calculate and compare the Deviance Information Criterion (DIC, \cite{refDIC}), the Akaike Information Criterion (AIC, \cite{refAIC}), and the Bayesian Information Criterion (BIC, \cite{refBIC}).  In the MRH package, there are two methods for obtaining these values.  The fitted model contains the different information criteria values (under the ``DIC", ``AIC", and ``BIC" labels), or the \makefxn{DIC} function that can be used on the fitted model or on the chains (see Section \ref{sec:usechains} for details on using the text files of chains).  Code for obtaining these values is below, and a comparison of the values for all models shown in this manuscript can be seen in Table \ref{tab:IC}. 
\begin{verbatim}
fit.NPHprune1$DIC
[1]  607.7004
fit.NPHprune1$AIC
[1] 667.0951
fit.NPHprune1$BIC
[1]  724.2637

DIC(fit.NPHprune1, n = 80)
$neg2loglik.summ
        value
Min.    585.8
1st Qu. 593.2
Median  596.0
Mean    596.5
3rd Qu. 599.3
Max.    619.1

$ICtable
       value
DIC 607.7004
AIC 667.0951
BIC 724.2637
\end{verbatim}
\begin{table}
	\caption{Comparison of model information criteria values (DIC, AIC, and BIC) for each of the models presented in the previous sections.  All values are calculated using the \makefxn{DIC} function in the MRH package.}\label{tab:IC}
\centering
\begin{tabular}{|c|l|l|ccc|}
	\hline
	\multicolumn{3}{|c|}{Model}&DIC&AIC&BIC\\
	\hline\hline
	PH&\multicolumn{2}{|l|}{Default}&614&676&721\\
	\hline\hline
	\multirow{7}{*}{NPH}&\multicolumn{2}{|l|}{Default}&613&697&783\\
	\cline{2-6}
	&\multirow{2}{*}{Pruned}&1 level&608&667&724\\
	&&3 levels&604&642&671\\
	\cline{2-6}
	&\multirow{3}{*}{Change $k$}&$k$ sampled&614&701&791\\
	&&$k = 0.1$&614&705&790\\
	&&$k = 10$&606&690&776\\
	\cline{2-6}
	&\multicolumn{2}{|l|}{Sample $\gamma$}&613&765&922\\
	\hline
\end{tabular}
\end{table}
\subsection{Settings and convergence diagnostics of the MCMC chains}\label{sec:MCMCconvergence}
The MRH package allows the user to control different properties of the MCMC chain (i.e. the burn-in, the thinning value, and the maximum number of iterations).  In addition, the \makefxn{estimateMRH} routine checks for possible convergence, and can modify the properties of the MCMC chain.  Plots are also provided in the output folder that allow the user to assess if the chains have converged.  Below, we outline the different methods for fixing the properties of the MCMC chain and assessing convergence.

\subsubsection{Setting parameters of the MCMC sampling}
The user may use the default values for the burn-in (``burnIn"), thinning value (``thin"), and maximum number of iterations (``maxIter") of the MCMC chains, or they may specify and fix these values. The default maximum number of MCMC iterations is set at 500,000, with a burn-in value of 50,000 and a thinning value of 10.  However, based on evidence of chain convergence, these numbers may be changed by the routine unless otherwise specified by the user.

\subsubsection*{Checking for evidence of convergence}

\noindent After the first 100,000 MCMC iterations, the chains are checked for autocorrelation and evidence of convergence.  (If the maximum number of iterations specified by the user is less than 100,000, then convergence is checked when the maximum number has been reached.)  The convergence checking routine is performed via the Geweke diagnostic test (\cite{geweke}) and the Heidelberger-Welch diagnostic test (\cite{HWelch}) using the \makefxn{geweke.diag} and \makefxn{heidel.diag} functions available in the coda package (\cite{coda}).  The convergence algorithm can be seen in the \ref{alg:convergence} Algorithm table.
\begin{varalgorithm}{Convergence}
\caption{Algorithm for auto-correlation and convergence checking}
\caption{Determine if convergence has been reached and auto-correlation minimized}
\label{alg:convergence}
\begin{algorithmic} 
\STATE
\WHILE{The number of iterations is less than \texttt{maxIter}}
\STATE
\STATE 1. Check if thinning value is high enough to reduce autocorrelation: Use \makefxn{acm} in the coda package to test the current value of \texttt{thin} as well as 5, 10 and 15 times \texttt{thin}.  The first lag where autocorrelation disappears is the new thinning value.  (Note that this value can only be greater than the previous value.)
\IF{New thinning value is different than original value}
\STATE Set \texttt{thin} to the updated value.
\STATE Thin the stored chains.
\ENDIF
\STATE
\STATE 2. Convergence check: calculate all p-values for the z-scores returned for each parameter from the \makefxn{geweke.diag} function and calculate the results of the Heidelberg-Welch test.
\IF{All p-values $< 0.005$ and the Heidelberg-Welch test passes}
\STATE Convergence reached.  
\ELSE
\WHILE{Convergence is not reached and the number of retained iterations (after thinning) is greater than 1000}
\STATE Burn 20,000 more iterations
\STATE Perform convergence check again. 
\ENDWHILE
\ENDIF
\STATE
\IF{Convergence is reached}
\STATE End MCMC sampling routine.
\STATE Update the values of \texttt{burnIn} and \texttt{thin} (if needed).
\ELSE
\STATE Set \texttt{thin}, \texttt{burnIn}, and stored chains to original values (from before Step \#1).
\STATE Perform another 100,000 MCMC iterations. (Or the remaining MCMC iterations needed for \texttt{maxIter} to be reached, whichever is smaller.)
\ENDIF

\STATE
\ENDWHILE
\end{algorithmic}
\end{varalgorithm}

There may be instances in which the user wants to fix \texttt{burnIn}, \texttt{thin}, or \texttt{maxIter} so that the routine does not change these values in the process of checking for convergence, and so that the user is guaranteed the MCMC chain will run \texttt{maxIter} times.  These values can be fixed (simultaneously or individually) by setting the ``fix.burnIn", ``fix.thin", or ``fix.max" options to TRUE.  (By default, these are set to FALSE.)

\begin{figure}[h!]
	\centering
	\includegraphics[width = 4.5in]{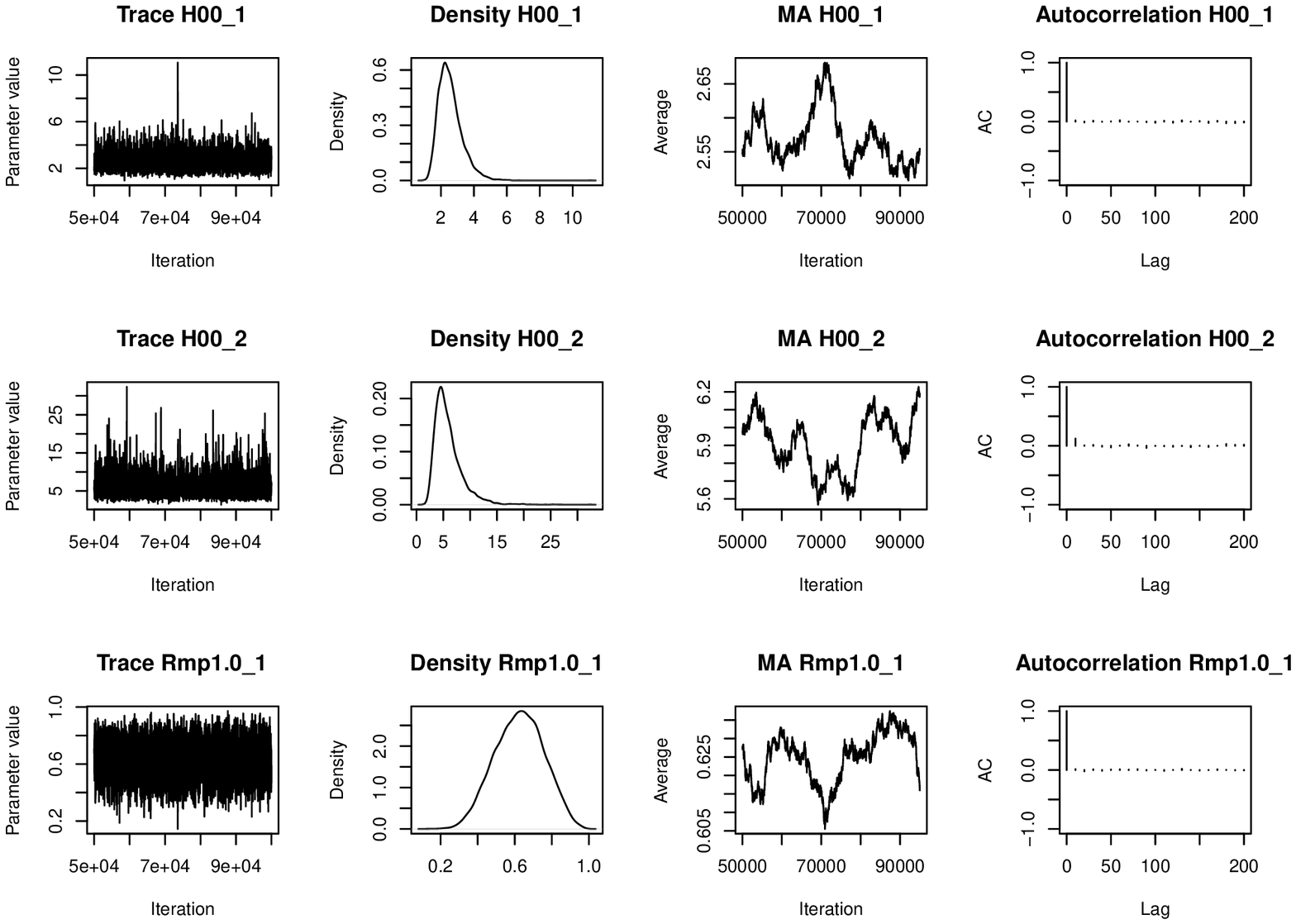}
	\caption{\footnotesize Trace, density, moving average (calculated in increments of 100), and autocorrelation plots for three of the parameters of the NPH 3-level pruned model.  (This graph can be found in the ``MRHresults\_NPH\_prune3" output folder, and is titled ``convergenceGraphs1.pdf".)  This graphic allows to user to also assess if convergence has been reached, or if certain parameters have a more robust estimate than others.}
	\label{fig:convplots}
\end{figure}

In addition to the convergence checking performed by the algorithm, diagnostic figures containing trace, density, moving average (calculated by 100), and autocorrelation plots for each parameter are included in the output folder for the model.  An example of these plots (for three parameters) can be seen in Figure \ref{fig:convplots}. This graphic allows to user to also assess if convergence has been reached, or if certain parameters are more problematic then others.

\subsubsection*{Continuing chains}
If the routine does not detect evidence of convergence and \texttt{maxIter} is reached, then the following message is shown to the user:
\begin{verbatim}
Warning message:
In estimateMRH(Surv(time, delta) ~ nph(type), data = tongue, M = 4,  :
  Algorithm has not yet converged after  MCMC iterations. 
					  Parameter estimates may not be reliable. 
\end{verbatim}
In this instance, the user may want to continue running the MCMC routine using the previously sampled chain values.  This can be done using the same call as the in the first model and by setting ``continue.chain" equal to TRUE.  (Note that this will only work if the output folder name is the same as with the previous model.  In addition, if the user specifies new thinning or burn-in values, these will be ignored -- the values from the previous chain will be used instead.)  If the chain is continued, the routine reads in the chains from the previous set of iterations and initializes the parameters using the last line of retained MCMC values, appending new samples to the previous existing text file store in the output folder. Example code and output for this is below:
\begin{verbatim}
fit.NPH.continue = estimateMRH(Surv(time, delta) ~ nph(type), data = tongue,
  M = 4, maxStudyTime = 400, outfolder = 'MRHresults_NPH_continue', maxIter = 5000)
....
Warning message:
In estimateMRH(Surv(time, delta) ~ nph(type), data = tongue  :
  Algorithm has not yet converged after  MCMC iterations. 
					  Parameter estimates may not be reliable. 

fit.NPH.continue = estimateMRH(Surv(time, delta) ~ nph(type), data = tongue,
  M = 4, maxStudyTime = 400, outfolder = 'MRHresults_NPH_continue', maxIter = 50000)

 MCMC routine running.  Calculating estimated runtime for 50000 iterations... 

Estimated total run time is 20 minutes 

To shorten the run time, re-run with fewer iterations or a smaller number of bins. 
\end{verbatim}

\subsubsection{Gelman-Rubin diagnostic testing}
To provide robust estimates of the hazard rate and covariate effects, it is common to run the MCMC routine for the same model multiple times, using the results from all chains to produce estimates and to check for convergence.  The MRH package provides a number of options that allow the user to do this more easily.  

In the \makefxn{estimateMRH} routine, the user can set  the ``GR'' (Gelman-Rubin) option equal to TRUE (by default this is FALSE).  In doing this, the routine automatically fixes the chain parameters (\texttt{thin}, \texttt{burnIn}, and \texttt{maxIter}) to what has been specified by the user in the function call, and also adjusts the initial values of the parameters to cover the parameter space.  (These are necessary qualifications for the use of the Gelman-Rubin convergence test.)

After all MCMC chains have been sampled, \makefxn{AnalyzeMultiple} is available for analyzing multiple MCMC chains for the same model.  The \makefxn{AnalyzeMultiple} function accepts multiple chains as an input parameter, and then returns to the user the estimated parameter values and $\alpha$-level credible intervals and the Gelman-Rubin diagnostic information.  Within each set of chains, the median, $\alpha/2$\%-tile, and $1-\alpha/2$\%-tile of the marginal posterior distribution is calculated for each parameter.  Then, the median of the medians and percentiles is calculated, and these are the numbers reported as the estimates and credible intervals for each parameter.  Example code for this is below:

\begin{verbatim}
# Fit the three models separately
fit.NPHprune31 = estimateMRH(Surv(time, delta) ~ nph(type), data = tongue,
      M = 4, maxStudyTime = 400, outfolder = 'MRHresults_NPHprune31', GR = TRUE,
	prune = TRUE, prune.levels = 3)

fit.NPHprune32 = estimateMRH(Surv(time, delta) ~ nph(type), data = tongue,
      M = 4, maxStudyTime = 400, outfolder = 'MRHresults_NPHprune31', GR = TRUE,
	prune = TRUE, prune.levels = 3)

fit.NPHprune33 = estimateMRH(Surv(time, delta) ~ nph(type), data = tongue,
      M = 4, maxStudyTime = 400, outfolder = 'MRHresults_NPHprune31', GR = TRUE,
	prune = TRUE, prune.levels = 3)

\end{verbatim}
An examination of the initial starting values shows that the initialized values cover the parameter space (which can be accessed with \texttt{fit.NPH1\$initialValues}).
\begin{verbatim}

# Get the parameter estimates and credible intervals using all three chains.  
# Also check for convergence using the Gelman-Rubin diagnostic test.
results = AnalyzeMultiple(fileNames = c('MRHresults_NPHprune31/MCMCchains.txt',
      'MRHresults_NPHprune32/MCMCchains.txt',
      'MRHresults_NPHprune33/MCMCchains.txt'), maxStudyTime = 400)
names(results)
[1] "hazardRate"       "beta"             "SurvivalCurve"    "CumulativeHazard"
[5] "d"                "H"                "Rmp"              "gelman.rubin"    
results$gelman.rubin
          Scale Reduction Factor
H00_1                         1
H00_2                         1
Rmp1.0_1                      1
Rmp4.2_1                      1
Rmp4.3_1                      1
Rmp1.0_2                      1
Rmp4.1_2                      1
Rmp4.3_2                      1
a_1                           1
a_2                           1
lambda_1                      1
lambda_2                      1
\end{verbatim}
The results of the Gelman-Rubin diagnostic test can be used to determine if convergence has been reached.  If the scale reduction factor is ``far from" 1, then it is recommended that more iterations be performed (using the ``continue.chain" option).

\subsection{Using the MCMC text files}\label{sec:usechains}
The computation time for convergence of the MRH model may be longer than the user would like to spend waiting (particularly in cases where the number of bins and/or the number of subjects is high).  In these instances, the user may want to run the model as a background job, in which case the fitted model results will not be available in the console.  Under these circumstances, the user can read in the MCMC chains from the output folder, convert them to an MRH object, and then may use existing functions for plotting and summarizing the chains.  The only difference between the usage of the functions is that the maximum study time (``maxStudyTime") must be entered for accurate estimates:
\begin{verbatim}
# Read in the file from the output folder
mcmc.NPH.prune3 = read.table('MRHresults_NPH_prune3/MCMCchains.txt', header = TRUE)

# Convert to an MRH object
 MRH.NPH.prune3 = as.MRH(mcmc.NPH.prune3)
class(MRH.NPH.prune3)
[1] "MRH"

#Summarize and plot the results (warning occurs if maximum study time is not entered)
summary(MRH.NPH.prune3)
Error in summary.MRH(MRH.NPH.prune3) : 
  Maximum study time (maxStudyTime) needed for hazard rate calculation. 
  The maximum study time can be found in the MCMCInfo.txt file in the output folder.
names(summary(MRH.NPH.prune3, maxStudyTime = 400))
[1] "hazardRate"       "beta"             "SurvivalCurve"    "CumulativeHazard"
[5] "d"                "H"                "Rmp"             

plot(MRH.NPH.prune3, maxStudyTime = 400)

\end{verbatim}
\section{Discussion}

In this manuscript, we have highlighted the main features of MRH package and demonstrated its use on the tongue cancer data set.  Use of the pruning tool and the accommodation of non-proportional hazards makes this package idea for estimation of right-censored survival outcomes when the number of observed failures is small and/or the follow-up period is long.

There are a few other packages that provide a non- or semi-parametric estimate of the hazard rate, and that accommodate non-proportional hazards. Namely, these are \texttt{bayesSurv} (\cite{bayessurvR}), \texttt{DPpackage} (\cite{dppkgR}), and \texttt{timereg} (\cite{timeregPkg}).  While these packages provide their own unique strengths to the analysis of right-censored survival data, the covariate interpretations for all three models are different than that of the MRH model: \texttt{bayesSurv} implements AFT survival models, \makefxn{LDDPsurvival} in the \texttt{DPpackage} package estimates covariates in an ANOVA-like fashion using a Dirichlet Process prior, and the \makefxn{timecox} function in the \texttt{timereg} package produces cumulative covariate estimates.  (For a thorough comparison of these packages, see \cite{RuserCompare}.)  The MRH package provides a useful tool for estimation of the hazard rate and covariate effects.

\section{Acknowledgements}
This work was supported by grants NSF-GEO 1211668 and NSF-DEB 1316334.  The project utilized the Janus supercomputer, which is supported by the National Science Foundation (award number CNS-0821794) and the University of Colorado - Boulder. The Janus supercomputer is a joint effort of the University of Colorado - Boulder, the University of Colorado - Denver, and the National Center for Atmospheric Research. Janus is operated by the University of Colorado - Boulder.  The authors thank Yuanting Chen and the researchers at NCAR and the REACCTING project for their helpful advice and input.

\end{document}